# Faster-is-slower effect in escaping ants revisited: Ants do not behave like humans

D. R. Parisi[*a, b], S. A. Soria[c], R. Josens[a, c]

a - Consejo Nacional de Investigaciones Científicas y Técnicas, Argentina.
b - Instituto Tecnológico de Buenos Aires, 25 de Mayo 444, Piso 5º, (C1002ABJ), C. A. de Buenos Aires, Argentina.
c - Grupo de Estudio de Insectos Sociales, Dto. de Biodiversidad y Biología Experimental, Facultad de Ciencias Exactas y Naturales, Universidad de Buenos Aires, Pabellón II, Ciudad Universitaria (C1428EHA), C. A. de Buenos Aires, Argentina.

**Abstract**

In this work we studied the trajectories, velocities and densities of ants when egressing under controlled levels of stress produced by a chemical repellent at different concentrations. We found that, unlike other animals escaping under life-and-death conditions and pedestrian simulations, ants do not produce a higher density zone near the exit door. Instead, ants are uniformly distributed over the available space allowing for efficient evacuations. Consequently, the faster-is-slower effect observed in ants (Soria et al., 2012) is clearly of a different nature to that predicted by de social force model. In the case of ants, the minimum evacuation time is correlated with the lower probability of taking backward steps. Thus, as biological model ants have important differences that make their use inadvisable for the design of human facilities.

**Keywords**: emergency; evacuation; egress; ant egress; crowd egress; faster is slower; faster is faster; pedestrian evacuation; pedestrian dynamics.

# 1. Introduction

It is already known that under normal conditions, ant and human fundamental diagrams (FD) are different (John et al., 2009). While the former displays a constant velocity for all densities, the pedestrian and vehicular FDs always show a monotonically decreasing velocity for increasing density (see for example Seyfried et al., 2005). Moreover, ants do not produce jamming (Dussutour et al., 2004; John et al., 2009).

Contrary to what the title of Soria et al.'s paper (2012) would suggest, the authors claimed that there are differences between ants and humans in highly competitive situations such as emergency evacuation through narrow exits. They reported the observation of the faster-is-slower (FIS) effect in escaping ants stressed with a chemical repellent at different concentrations. Even though the FIS effect has been reported for simulated pedestrians via de social force model (SFM) (Helbing et al., 2000), it is not enough for justifying an analogy between ants and humans when egressing through a narrow door. One should not be misled by the title of the paper since the authors clearly state that the mechanisms causing the FIS effect in ants are not the same as those in the SFM simulations. So, although the "FIS effect" was reported in this paper, using it as a proof that ant and human egress is similar, it is not correct.

During different types of emergencies people can adopt different behaviors depending on the demand and capacity of the means of egress. The balance between the demand and capacity is given by several factors such as the kind of physical threat, information and subjective perception of danger, the number of people and the widths of the means of egress. As long as the physical threat is not imminent or not directly perceived (for example an alarm, but no smoke or fire), people tend to be cooperative (Kretz, 2010). The shorter the time (or the smaller the width of a door or stairway) available for a safe egress is, the lower the degree of cooperativeness results. As there is less time available for escape from danger, decisions under stress could be taken (Keinan et al., 1987), which could result, for example, in choosing the main entrance instead of the nearest exit as means of egress. In the extreme case that the time available is very scarce to escape from a sure death, the predominant behavior would be the individual self-preservation.

In such a situation, people could choose rushing or not rushing toward the exit. As this decision has an impact on the payoffs of each agent and the whole group, it can be studied from the point of view of game theory. Heliövaara et al. (2013) have shown that jamming and clogging may be caused by people acting rationally, even when this rational individual behavior results in a bad strategy for the group.

An example of such egressing behavior, which saturates the capacity of the egress door, is the fire at "The Station Night Club" (Rhode Island, USA, 20 February 2003), where an amateur camera recorded the tragedy (http://www.youtube.com/watch?v=OOzfq9Egxeo). There, it can be seen that, at a given moment, most of the people tried to egress simultaneously through the main door, causing the blockage of that door (Fahy et al., 2012).

The behavior of rushing toward a door was also observed in animals in less frightening situations. Saloma et al. (2003) found this response when studying the egress of mice from a water pool. Zuriguel et al. (2014) also observed the same response when studying the anxious passage of sheep through a narrow door when they were to be fed.

Also, this selfish evacuation behavior is the one assumed in the paper where the FIS effect was first reported (Helbing et al., 2000). The cause of this effect is the high tangential friction between particles in contact (Parisi and Dorso, 2007). Moreover, the FIS effect was recently verified experimentally in granular media (Gago et al., 2013), herd of sheep (Zuriguel et al., 2014), and humans (Garcimartín et al., 2014). In all cases the FIS effect appears due to jamming and clogging at the door, producing high frictional forces.

On the other hand, the FIS effect reported for ants (Soria et al., 2012) was not caused by any kind of frictional contact, jamming or clogging. This behavior was also confirmed in another experiment with Argentinean ants stressed with temperature (Boari et al., 2013) in which, contrary to the FIS effect, the "faster-is-faster" effect was found, even when ants were close to dying by temperature (if it had risen a little bit further).

The present work is based on the video recording from the experiments performed in Soria et al. (2012). Here we used image processing technics for obtaining the individual trajectory of ants. From this information, velocities and densities can also be studied. These data allowed us to demonstrate the claim that ants do not jam nor clog near the exit and thus that the FIS observed has no relation to the FIS effect in other animals' systems. To be even more specific, we are going to compare the more relevant metrics obtained from ant data with the corresponding ones from simulations with the SFM producing the real FIS effect. As a consequence, it will be evident that the FIS effect in ants is not the same as in other systems relevant to the area of highly competitive egress.

## 2. Materials and methods

We analyzed the recorded video of the experiments reported in Soria et al. (2012) when studying the egress of *Camponotus mus* (*Roger*) ants stressed with aversive stimuli through a narrow exit. A detailed description of the experiments can be found in that paper; here we only summarize the main features.

Approximately one hundred ants were placed in a transparent arena consisting of a floor, walls and a ceiling so high that ants could not get on one another, thus maintaining the system two-dimensional.

In order to produce different degrees of repellency, ants were exposed to different concentrations of a repellent solution, made with water and citronella. The increasing degree of repellency mimics the increase of the desired velocity in the SFM simulation of the room evacuation problem (Helbing et al., 2000), where the FIS effect is observed. The repellent was placed at the wall opposite the door.

The geometry of the arena can be seen in Fig.1 A and it was designed to guide ants toward the door. Before the egress began, ants were allowed to enter the punishable chamber using the only exit from the non-punishable chamber. At the start of the evacuation, the citronella solution was placed and the door was opened simultaneously. Trials were recorded with a HD camera at 30 fps.

Image processing of the video recording was performed using the tool built by Liendro and Goldberg (2013), which allows one to obtain the position of each ant as a function of time.

As one of the main objectives of this work is to show that the FIS effect observed in ants has a different nature to that arising from the social force model (SFM) (Helbing et al., 2000), we wanted to compare the metrics obtained for ants with the corresponding ones obtained from simulations with the SFM in the same geometry. An equivalent arena was considered (Fig.1 B) scaling its dimensions for using particles corresponding to the pedestrian simulations. The radius of particles ranges from 0.23 to 0.25 m. The criteria used is that the exit width is about 2 particle diameters, similar to the ant experiment where the exit width is about 2 ant widths. The pedestrian mass was uniformly distributed between 70 and 90 kg. We performed simulations of this system with desired velocities $v_d=0.6; 1; 2; 4$ and $6$ m/s. The model and the rest of the parameters were the same as in Helbing et al. (2000).

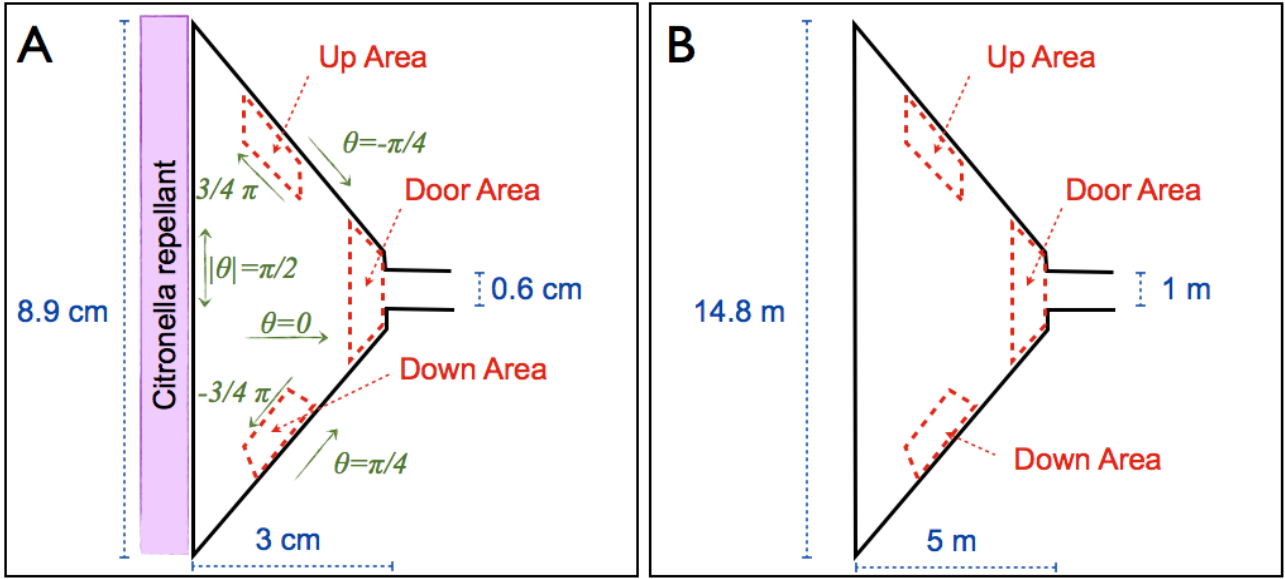

**Figure 1.** Geometry for the evacuation process and the 3 areas for density measurements. A) The punishable chamber of the arena used in the ant experiment (Soria et al., 2012). Definition of the direction angle of the velocity ($\theta$) for individual ants. B) Scaled geometry for simulating pedestrian egress using the social force model (Helbing et al., 2000).

*2.1 Definition of metrics*

In this section we will define some observables derived from the trajectories of the agents, either ants or simulated pedestrians.

*2.1.1 Time lapses between two outgoing ants*

By registering the time at which each ant goes out from the punishable chamber, the curve of the number of exited ants vs. time can be built. Some examples of these curves are displayed in Fig. 2 A. From these data the time elapsed between two successive outgoing ants (*dt*) for the first 70 ants was measured in all the trials at different concentrations.

*2.1.2 Ant velocities*

Let us consider the trajectory of any particular ant given by the vectorial function $\mathbf{x}(t_i)$, where $\mathbf{x}$ is the position at the plane of the arena and $t_i$ is the sampling time of the video recording at 30 fps, meaning that $(t_{i+1} - t_i) = 1/30$ s for all $i$. At each discrete time $t_i$ the velocity of this particular ant can be obtained through a discrete difference as $\mathbf{v}(t_i) = [\mathbf{x}(t_{i+k}) - \mathbf{x}(t_i)] / (t_{i+k} - t_i)$. The parameter $k$ is the number of time steps taken to obtain the velocity. The better approximation to an "instant" velocity for these data is $k=1$. However, in order to avoid high-frequency fluctuations of the experimental trajectories, we computed the velocity over 3 time steps by setting $k=3$.

The velocity vectors $\mathbf{v}(t_i)$ could also be represented by their magnitude and angle. We will call this angle $\theta_i$ and it is defined such that $\theta_i = 0$ coincides with the horizontal direction from left to right, as shown in Fig. 1 A. Other relevant values of $\theta$ are also displayed in the same figure.

We will also focus our attention on the changes of direction in each ant trajectory for all discrete times $t_i$ that are given by $d\theta_i = (\theta_{i+1} - \theta_i)$.

*2.1.3 Densities for ants and simulated human systems*

From the trajectories $\mathbf{x}(t_i)$, either from an ant experiment or pedestrian simulations, the density at any given area is computed by counting the number of agents inside of it and dividing by the corresponding area: Density $\rho_j(t_i) = N_j(t_i)/A_j$, where $N_j(t_i)$ is the number of agents at time $t_i$ in the measurement area $A_j$.

Note that as $N_j$ evolves when agents enter and go out from a given measurement area, this is a function of time and so, the density will also be a function of time.

With this methodology, the density can be calculated at different areas. By dividing the whole area of the arena into a square grid of smaller areas, a density map can be obtained at any time step or by averaging over the whole evacuation process.

We considered a grid of 0.3 cm x 0.3 cm for the ant arena and a grid of 0.5 m x 0.5 m in the case of the pedestrian system.

Also, we wanted to study the density at selected relevant measurement areas that are shown in Figs. 1 A and B and we called them: door area, up area and down area following the nomenclature in this figure. These areas are about 1.5 cm$^2$ in the case of the ant arena (Fig. 1 A) and about 2 m$^2$ in the case of pedestrian simulation (Fig. 1 B).

Taking into account these three selected areas, a final density metric considered was the "Door Density Ratio," defined as the average density at the door area divided by the maximum density of the average of the other two areas. If this ratio is greater than one, it indicates that the mean density at the door area is greater than the densities at the other areas.

## 3. Results: Ants experiments and pedestrian simulations

### *3.1 Distribution of time lapses "dt"*

In this section we will confirm the FIS effect, reported in Soria et al. (2012), using different metrics and study the distribution of *dt*.

By grouping the time lapses *dt* (see Sec. 2.1.1) according to the citronella concentration used in the trial, the complementary cumulative distribution function (CDF) of *dt* can be computed for the four distributions as shown in Fig. 2 B. The lowest distribution corresponds to 75% citronella concentration, confirming the reported result when the mean evacuation time for the first 70 ants was measured (Soria et al., 2012).

These distributions were analyzed using the methodology proposed by Clauset et al. (2009) and it turned out that they did not display a power-law tail (p-value < $10^{-2}$). This is very important because in recent observations of experimental FIS effects in granular material (Gago et al., 2013), herd of sheep (Zuriguel et al., 2014) and humans (Garcimartín et al., 2014) the tail of the distribution of *dt* was found to be power-law. This result indicates a very significant difference in the nature of egress in ants and other systems where the friction is relevant.

Given that the distributions of time lapses are nearly exponential, the first moment (the mean) of the distribution is well defined. In Fig. 2 C the means of *dt* are plotted against citronella concentration and a minimum can be observed at 75% concentration, which is consistent with the previous finding.

An alternative way of viewing this result is considering that the flow rate is the inverse of the mean time lapse, of course in this case the minimum time lapse corresponds to the maximum flow rate, which is shown in Fig. 2 D.

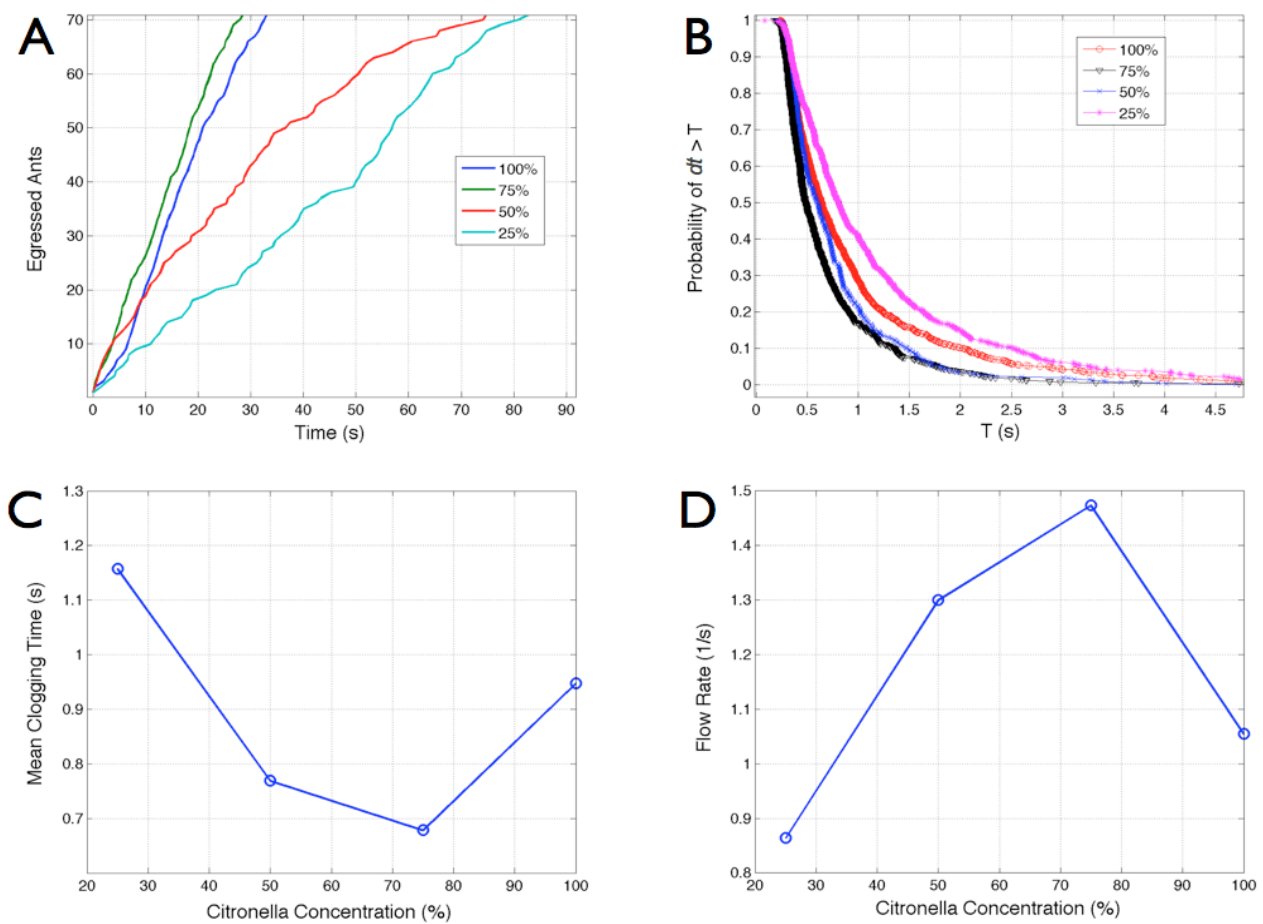

**Figure 2.** Ant experimental data. A) Examples of the discharge curve of four trials with the four citronella concentration used. B) Complementary cumulative distribution function of the time lapses between the egresses of two consecutive ants ($dt$). C) Average of $dt$ as a function of citronella concentration. D) Flow rate of outgoing ants as a function of citronella concentration.

## 3.2 Density comparison

In this section we will study the density in the whole area of the punishable chamber and in the areas defined in Fig. 1, for all the trials performed in the ant experiments.

Figure 3 shows snapshots of the evacuation process up to the first 70 agents along with the count of agents in each measurement area for both systems.

The difference between ants and simulated humans is evident. The agents in the simulations rush directly toward the exit causing the door area to remain saturated (Figs. 3 E and F), which produces jamming and clogging obstructing the exit and leaving the others areas (up and down) empty. On the contrary, ants look uniformly distributed in the punishable chamber, where the count at the up and down areas is similar to and sometimes higher than at the door area (Figs. 3 B and C). This behavior of ants was observed in all the trials independently of citronella concentration.

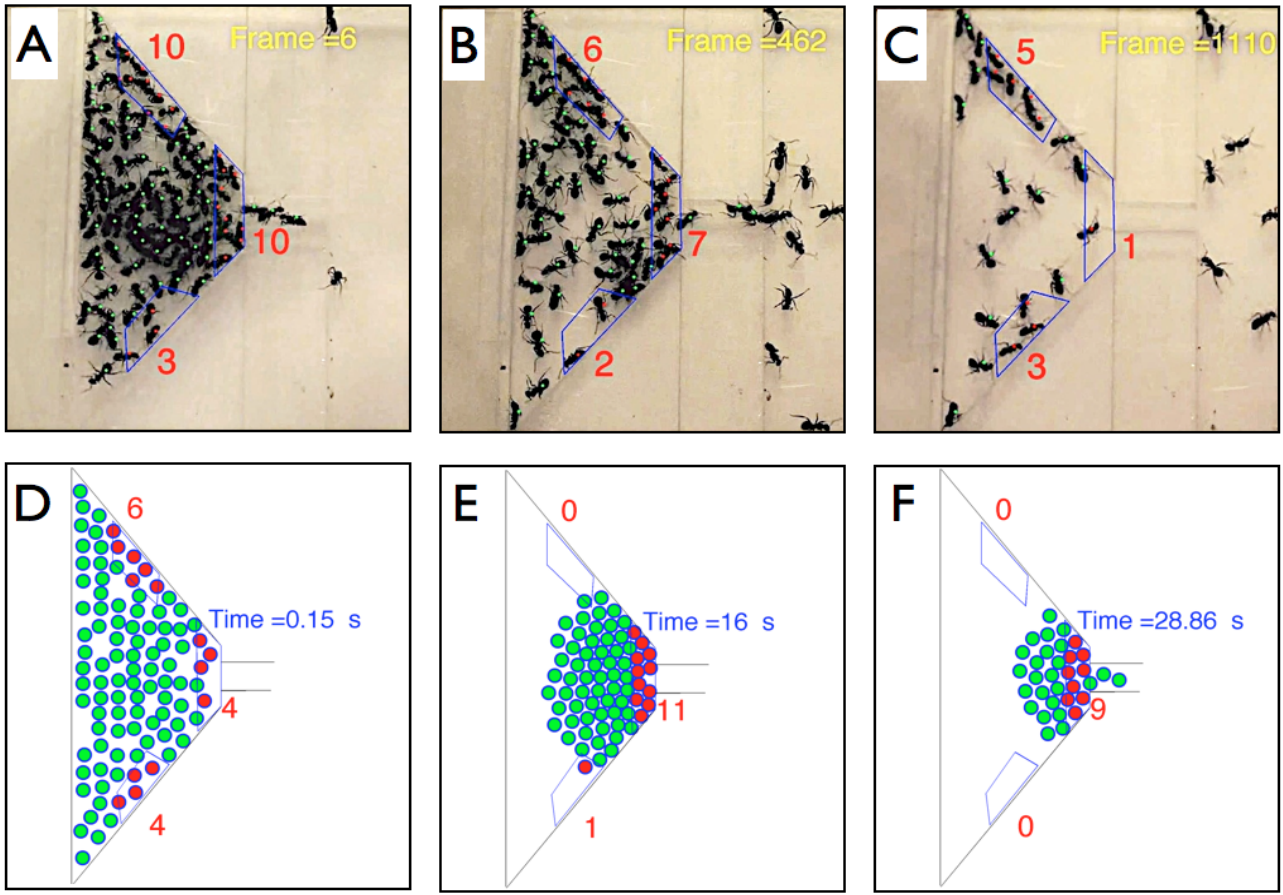

**Figure 3.** Differences between ant experiment and SFM simulations. Snapshots at different stages of typical evacuation processes. The egress of ants (in this particular case using 75% citronella concentration) is shown at the beginning (A), the middle (B) and at the end (C) of the evacuation. The evacuation process of pedestrians simulated using the SFM is shown at the beginning (D), the middle (E) and at the end (F) of the evacuation. In both cases the figures near the measurement areas are the number of agents in that area. Particles with red dots are inside some measurement areas, while the particles signaled with green are outside these areas.

As the density measurement can be made for each frame, the time evolution of the count can be obtained during the evacuation process. A running average of these signals was plotted for each system in Figs. 4 A and B. It can be observed in the case of ants (Fig. 4 A) that the evolution of the density of the three measurement areas was similar (within the fluctuations) and had a decreasing tendency according to the emptying process of the chamber and a uniform distribution of ants. In opposition to this pattern, simulated pedestrians (Fig.4 B) showed that the density at the door area saturated, after a short transient, and remained saturated during the whole process, while the other areas were emptying as the cluster of agents in front of the door decreased. This constant emptying of the up and down areas can be understood by looking at the frames shown in Figs. 3 E and F.
These continuous curves demonstrate that the snapshots shown in Fig. 3 are not random fluctuations but rather representative images of the evacuation process of each system.
But let us go further and consider all the density signals from all the trials grouped by citronella concentration, in the case of ants, or by desired velocity, in the case of simulated pedestrians. In order to do this we take, at each area, the average of all the time during the evacuation of the first 70 agents and of all the trials of a given citronella concentration. Figures 4 C and D display the results.

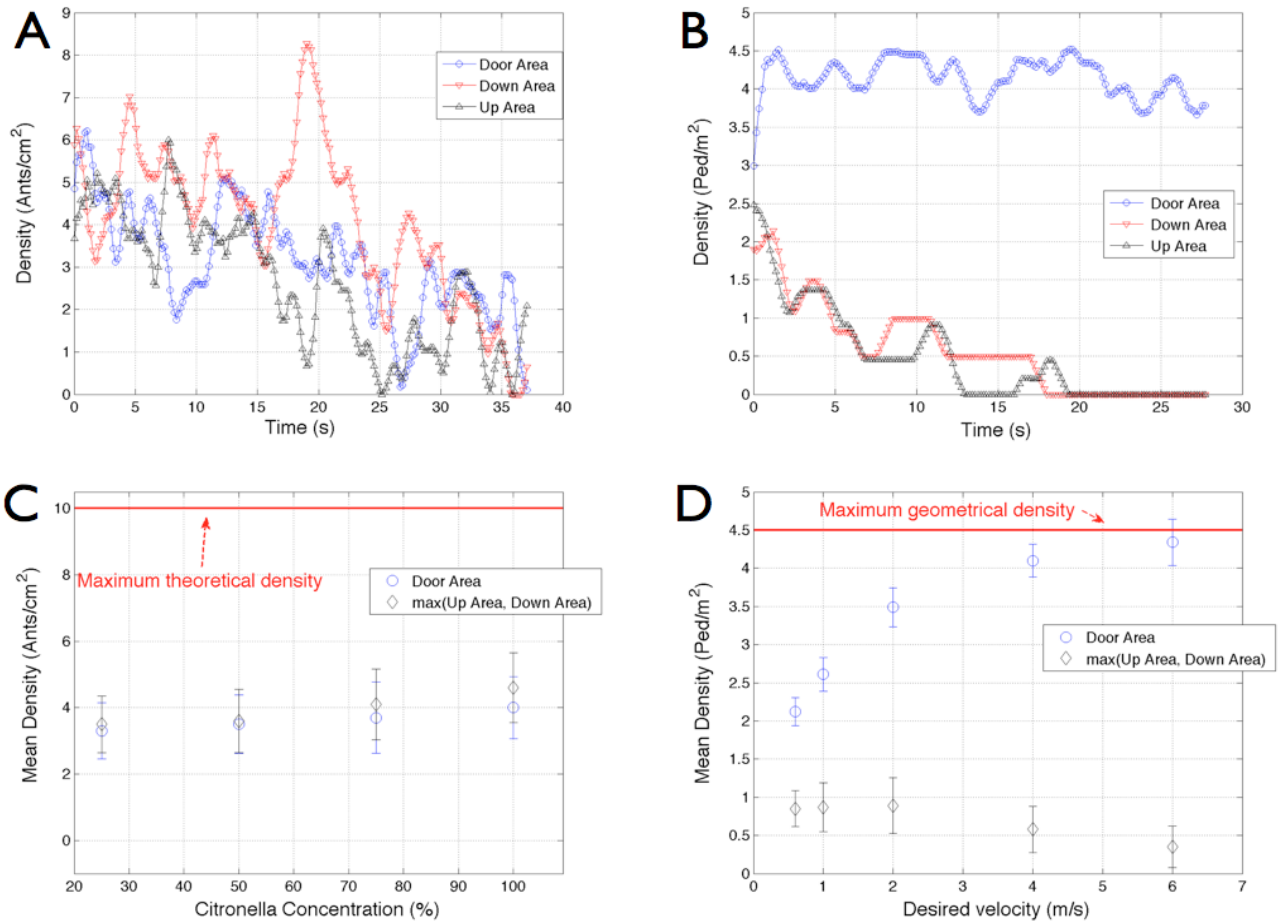

**Figure 4.** Differences between ant experiment and SFM simulations. Measurements of density at the areas given in Figs.1 and 3. A) Running average of the density signals for the three areas during the ant evacuation process with 75% citronella concentration. B) Idem for a simulation of the pedestrian system with $v_d$ = 4m/s. C) Mean value of the density at the door and the maximum density of the other two areas, for all the times and for all trials as a function of citronella concentration. D) Idem for the simulated pedestrian system but as a function of the desired velocity.

It can be seen that the differences in the evacuation process observed previously are consistent for all the ant experiments performed. The density at the door compared to the maximum density at any of the other areas (up or down) is very similar, even less. Also, the variations for different concentrations were very small and none of the density averages was close to the maximum possible density of ants.

At odds with the pedestrian simulation case, which shows a clear separation between the density at the door area and at the other two areas, the density at the door area was much greater and also increased with the desired velocity, tending to the value of maximum density reachable by this particle system.

The mismatch between ant and human egress can be even more evident if we plot the Door Density Ratio defined in Sec. 2.1.3 as a function of citronella concentration, in the case of ant experiments, and of the desired velocity ($v_d$), for the simulated pedestrian system. Figure 5 shows the clear difference between both systems; while the ratio is near 1 for all values of citronella concentration in the ant experiments, it increases by more than 12 times for $v_d$ = 6m/s in the pedestrian system.

As a final output of the density data in Fig.6 we present the density map for the complete area of the evacuating chamber divided into a grid of 0.3 cm x 0.3 cm for ant data and 0.5 m x 0.5 m in the pedestrian system.

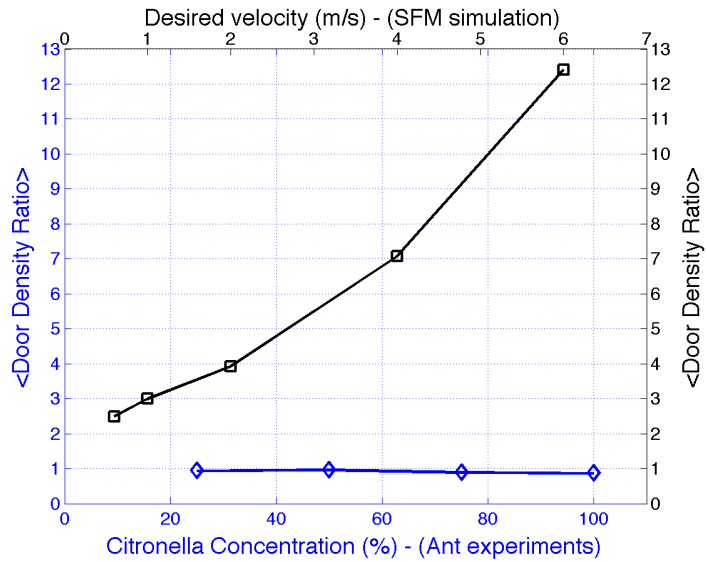

**Figure 5.** Comparison of the Door Density Ratio for ant experiments (rhombus, bottom X axis) and pedestrian simulations (square, top X axis) as a function of the degree of urgency (citronella concentration for ants and desired velocity for virtual pedestrian).

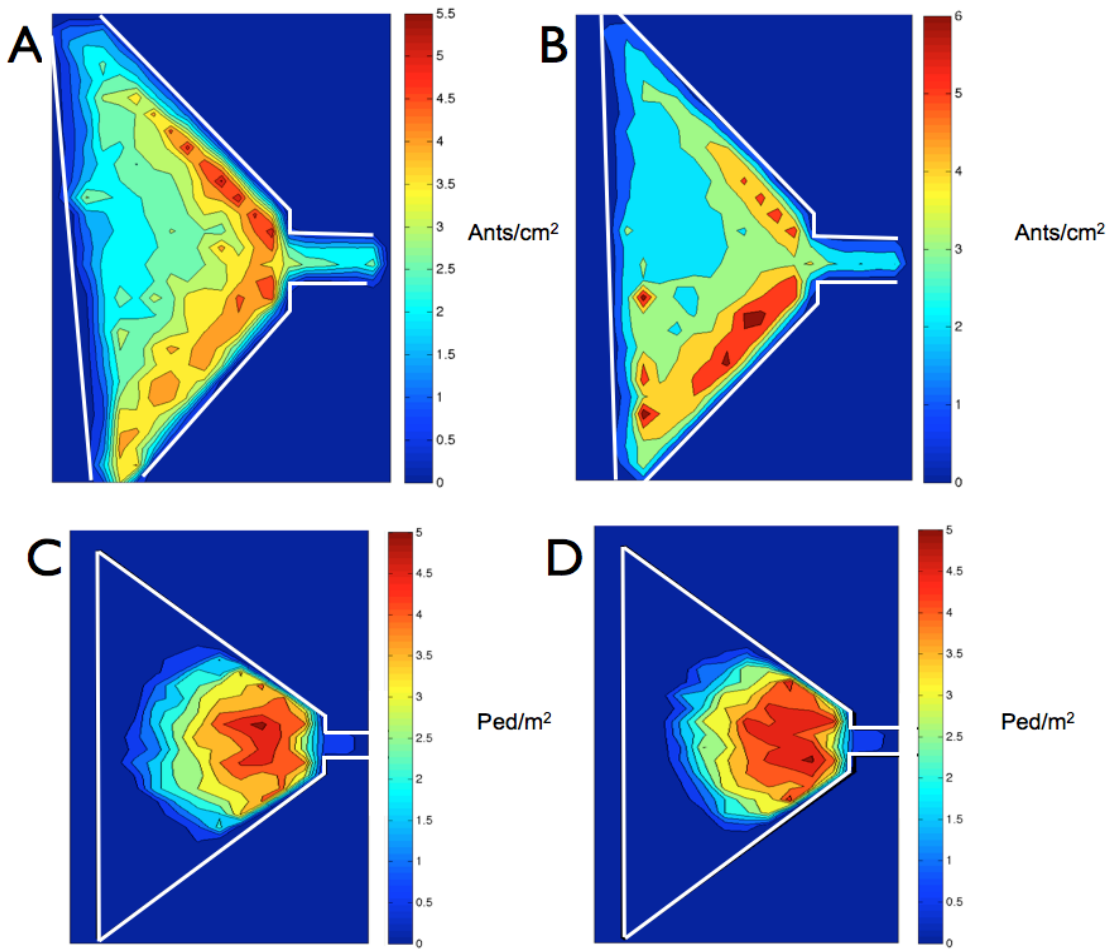

**Figure 6.** Comparison of mean density over the complete area for all the times and trials. A) For the ant experiments at 75% citronella concentration. B) Ant experiments at 100% citronella concentration. C) For pedestrian simulation with $v_d$=4m/s. D) For pedestrian simulation with $v_d$=6m/s.

Again, the differences are remarkable. For simulated pedestrians there was a high-density spot just in front of the door. While for ants, the access to the door remained free having a low-medium density. Ants used the whole area instead of conglomerating at the door area as in the case of the pedestrian simulation.

In the ant system, in the neighborhood of the wall where the repellent was placed, the lowest density values were observed, indicating that it delivered the information of "pushing" ants toward the door by means of concentration gradient. In other words, the presence of the repellent gave information inducing the direction of escape. Also, it can be observed that the maximum densities in the ant system were near the side walls, confirming that ants prefer walking near walls than into open spaces (Dussutour et al., 2005).

The density patterns observed in ant experiments do agree with the observations made, at normal conditions, by Burd (2006) and Dussutour et al. (2004) suggesting that ant traffic is ruled by cohesive and dispersive interactions, and apparently the dispersive forces may be responsible for the efficient ant traffic organization.

*3.3 Velocity analysis of ant experiment*

In this section we analyze the distribution of ant velocity directions ($\theta$) and changes of these directions ($d\theta$) as described in Sec. 2.1.2.

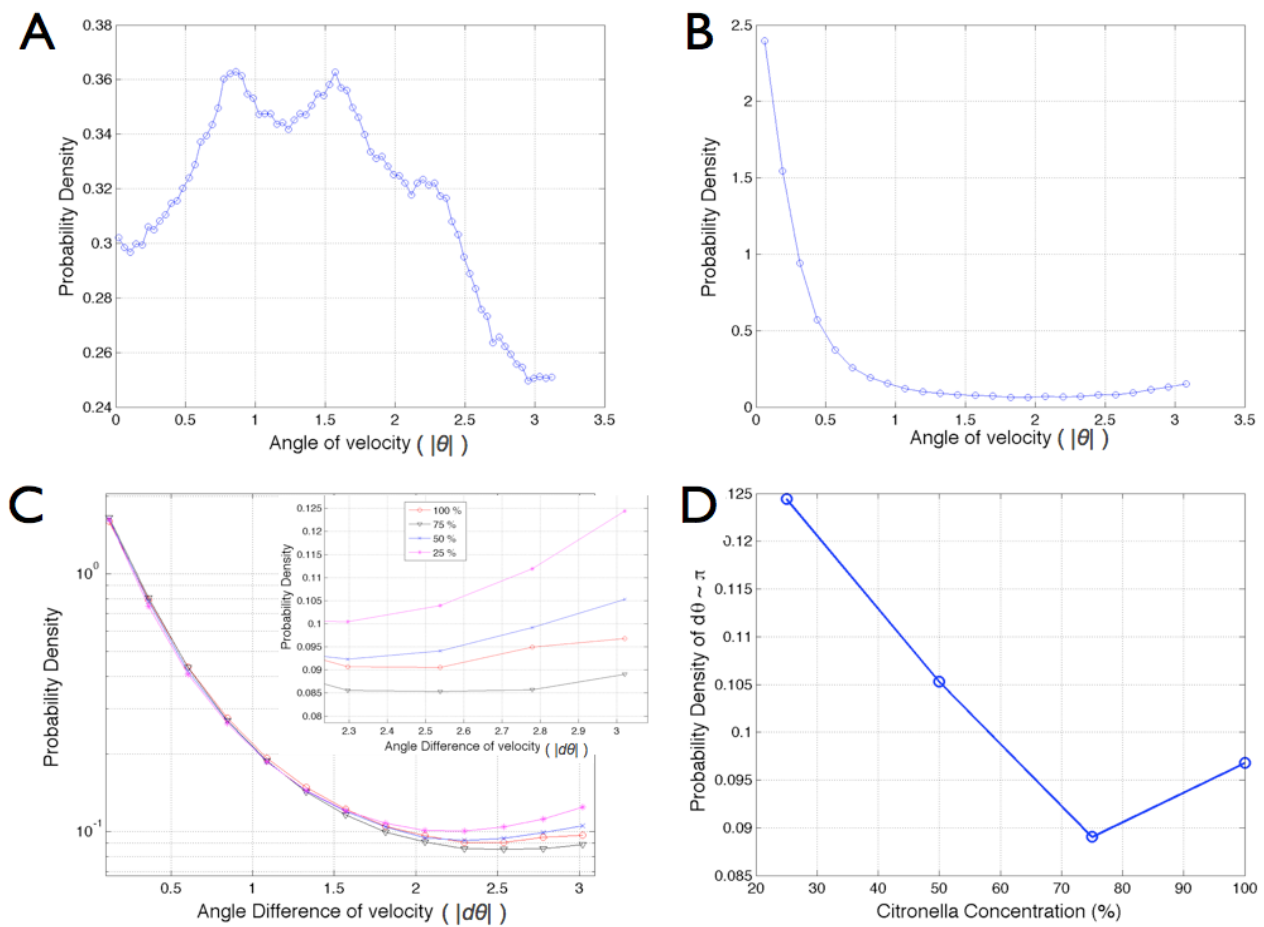

**Figure 7.** Metrics of the velocity directions. A) Probability density function of the absolute value of the velocity angle ($|\theta|$) inside the punishable chamber (Fig.1 A). B) Idem but in the corridor after the door connecting to the other chamber. C) Probability density function of the absolute values of changes in the velocity angle ($|d\theta|$) inside the punishable chamber. The inset displays a zoom in the zone of backward steps. D) Plot of the bin of $|d\theta| \sim pi$ as a function of citronella concentration.

The probability distribution functions of velocity directions (Fig.7 A and B) were very similar for all citronella concentrations and thus, the complete set of data was considered. Figure 7 A shows the distribution of $|\theta|$ inside the punishable chamber without considering the corridor after the door, which led to the other chamber. Two major peaks corresponding to $|\theta| = \pi/2$ and $\pi/4$ can be easily distinguished. These directions correspond to up and down along the repellent wall ($\theta = \pm \pi/2$) and along the lateral walls with the direction toward the door ($\theta = \pm \pi/4$) as defined in Fig. 1 A. These are the predominant directions followed by ants, which is further evidence of the tendency of ants to follow the walls.

The remarkable asymmetry between the direction $\theta = \pm \pi/4$ (toward the door along lateral walls) and $\theta = \pm 3/4 \pi$ (away from the door along lateral walls), where the former presented a greater probability of occurrence, is worth noting. This asymmetry shows the effect produced by the repellent and is consistent with the fact that there is a net flow of ants going out from the punishable chamber.

Another broken symmetry of the same type is the one between $\theta = 0$ (toward the door without following the walls) and $\theta = \pi$ (away from the door without following the walls). This also resulted in a net flow of ants toward the door.

The distribution of velocity angles in the corridor after the door is shown in Fig.7 (B). In this location there is a clear predominant velocity on the way to the exterior of the chamber ($\theta = 0$) limited by the geometrical constriction.

Differences between the four citronella concentrations studied can be observed if we look at the changes in the velocity angle $d\theta$ shown in Fig. 7 C. This time lag was taken to avoid noisy fluctuations due to the identification process of each ant center of mass. Processive (or straight-line) trajectories were characterized by $d\theta \sim 0$. This was the most probable case, meaning that trajectories were not random walks, but that they had a preferred direction during several time lapses. The probability of change in direction decreased for increasing $d\theta$ up to the neighborhood of $d\theta \sim \pi$, where this probability increases again, meaning an increase of proportions in backward steps taken. In this region, the data from different citronella concentrations were distinguishable and they showed the same tendency as in the FIS effect considering the means of time lapses $dt$ or the evacuation times. As can be observed in Fig. 7 D, 25% citronella concentration is the most probable case of taken backward steps, followed by 50% and 100%, with 75% citronella concentration being the one with less probability of $d\theta \sim \pi$.

So, looking at citronella concentration, the minimum of backward steps coincided with the minimum evacuation time. This result can be interpreted in terms of ant trajectories, which were more direct when they achieved maximum evacuation performance. For 100% citronella concentration, the trajectory changes go back to values of less urgency to escape. This can probably be caused by incoordination due to the influence of very high citronella concentration affecting the sensory and motor systems of ants.

## 4. Discussion

The fact shown in the present work and in Boari et al. (2013) that ants do not rush toward the door in a selfish evacuation behavior when stressed with citronella or heat (endangering their own individual life) can be because social insects (ants and honeybees, among others) are not expected to behave similarly to other animals because they have a different life cycle than that of individual animals (including humans), as the reproductive unit is not the individual but the colony. Ant societies, even when composed of thousands of individuals, behave coordinately as a whole (Detrain and Deneubourg, 2008; Fourcassié et al., 2010; Burd, 2006) and they should be expected to act more cooperatively than the rest of the animals, even under threatening conditions.

Independently of the causes of this deep difference between ants and other animals, the data presented are conclusive: ants do not jam nor clog near the door if they have available space.

This result shows that the methodology of using ants for studying pedestrian behavior under emergencies (Burd 2006; Shiwakoti et al., 2014 and previous work from the same authors referenced therein) should not be applied. For this reason, we strongly recommend not using experimental data from ants either for validating computer models of pedestrian dynamics or for designing human egress systems because even when humans can evacuate in a cooperative and ordered way during some kinds of emergency, they can also evacuate in a very competitive and selfish way as described in Sec. 1. Therefore, we propose instead that any other individual animal could be more similar to humans as biological model.

On the other hand, ant experiments could be used properly if we study what ants do under extreme conditions, because ant strategies maximize the performance of the whole group. Ants do egress efficiently, and we could use the knowledge obtained from their strategies to improve our egress protocols. A study on this topic has already been presented (Parisi and Josens 2013).

**5. Conclusions**

In the present work we have provided new data extracted from previous experiments by means of image processing technics. The data consisted of the positions of ants as a function of time (trajectories) during the egress under stressed conditions, which allowed us to study velocities and densities at different locations inside the arena.
The results from the ant experiments have shown that ants distributed uniformly over the available area during the egress process and did not generate high densities near the exit and thus, they performed efficient evacuations.
Comparison with an equivalent system of simulated pedestrians was made via the social force model (Helbing et al., 2000) assuming a selfish evacuation behavior that produced the faster-is-slower effect. The differences are remarkable. The ratio between mean density at the door area and at other inner areas was near 1 in the case of ants (according to a uniform distribution of ants over the arena) while for the case of simulated pedestrians it was as high as 12, meaning that the mean density at the door area was up to 12 times greater than in the other inner measurement areas.
Also, the time distribution of time lapses between the output of two consecutive particles was not power-law as in the case of recently experimental FIS measurements (Gago et al., 2013, Zuriguel et al., 2014, Garcimartín et al., 2014).
The evidence presented here further confirms the statement already claimed in Soria et al. (2012) and Boari et al. (2103) that the FIS effect observed in ants is not caused by contact, friction, jamming or clogging before the exit as in the case of pedestrians and other systems exhibiting the FIS effect.
The FIS effect observed in ants stressed with citronella is correlated with the probability of taking backward steps in the ant trajectories, probably caused by some kind of harmful effect on the physiology of ants due to the high citronella concentration.
In consequence, the FIS effect observed in ants is of a very different nature to that corresponding to pedestrian evacuation simulations and thus it cannot be used as a fact that allows the use of ants as a biological model of humans evacuating under emergency. As a matter of fact, it is the opposite; ants can behave very differently to the rest of the animals in life-and-death situations. This result will allow building more reliable computer models of emergency evacuations for pedestrians.

**Acknowledgments**

Roxana Josens and Daniel R. Parisi are Scientific Researcher at CONICET (Argentina). This work was supported by grant PICT 2011-1238 (ANPCyT, Argentina).